Tidal triggering of M7+ earthquakes by Jupiter

Elizabeth Holt, Eastern Sierra Unified School District and University of Arizona
ewholt3roripaugh@arizona.edu

Eric Newman, Oddr Inc.

ABSTRACT
This work uses a chi-squared test of independence to determine if days that include earthquakes greater than or equal to magnitude 7 (M7+) from 1960-2024 are truly independent of the position of Earth in its orbit around the sun. To this end, this study breaks up Earth's orbit into days offset on either side of two reference Earth-Sun-Planet orientations, or zero-points: opposition and inferior conjunction. A computer program is used to sample USGS earthquake and NASA Horizons ephemeris data for the last 64 years (1960-2024) with the purpose of conducting 28,782 chi-squared tests-of-independence for all intervals (5 to 45 days) spanning the entirety of Earth's synodic period relative to these reference points for Jupiter. 1,071 statistically significant intervals of non-random M7+ earthquake activity are associated with two particular points: the inferior conjunction and what is termed the "preceding neap" position. At both of these points, M7+ activity first increases (>125% of average) and then sharply decreases in a pulse-like fashion, with those lulls in M7+ activity (<75% of average) lasting about a month. Both of these pulses of M7+ activity begin at Sun-Observer-Target (SOT) Angles near 45 degrees and 135 degrees, and this is also observed for Venus and Saturn; Mars synodic period prevents any comparison of chi-squared intervals to SOT Angle. Although this study did not observe any obvious correlation of M7+ activity with the lunar cycle, the medians, means, and modes of the significant intervals returned by the chi-squared analysis for Jupiter range from 27 to 34 days, suggesting that intervals are more likely to be found significant by the chi-squared analysis if they average out the lunar cycle.

INTRODUCTION: The idea that the changing orientation of planets in the solar system might have some observable impact on Earth itself is a seductive and ancient concept developed by nearly every culture that observed the planets to move differently on the backdrop of the galaxy. Mythology, astrology, and a host of non-scientific theories regarding this subject abound, making it important to apply an increased level of mathematical vigilance when testing scientific hypotheses. Current theories regarding Earth's formation and early history do involve significant interactions with other bodies in the solar system: 1) the current well-regarded hypothesis for the formation of the moon involves collision with another planet-sized object (Kegerreis et al, 2022; Hartman and Davis, 1975; Cameron and Ward, 1976); 2) there is evidence that Earth's volatiles (water, carbon dioxide, etc.) originate from early-Earth impact events (Drake, 2005; Cowen, 2013; Mandt et al., 2024); and 3) occasionally, a near-Earth object makes its way through Earth's atmosphere to strike the ground. One of these larger objects is hypothesized to have triggered the extinction of the dinosaurs at the end of the

Cretaceous period (Alvarez, 1980). Further, it is well known that both the moon and the sun exert significant tidal forces on the Earth, measurably influencing the acceleration due to gravity (Hinderer and Crossley, 2000). In 1975, it was proposed that tidal forces trigger earthquakes on Earth (Heaton, 1975); however this work was retracted by Heaton in 1982, when he conducted a more sophisticated statistical analysis that revealed there was no consistent evidence for a correlation between tidal forces and earthquakes. Since then, many authors have brought forward anecdotal evidence to suggest that earthquakes, and in particular large earthquakes, are sometimes triggered by tidal forces [see Yan et al. (2023) for a recent comprehensive list].

The tidal forces the planets exert upon the Earth are so small compared to the tidal forces of the moon (at least 1000 times less for Venus and even smaller for the other planets) that it would appear highly unlikely that we should observe any correlation between seismic activity and planetary orientations, especially given that there is no clear consensus on whether the far stronger lunar tides consistently provoke seismic activity on Earth (Yan et al., 2023; Heaton, 1975 and 1982; Hartzell and Heaton 1989; Tanaka et al., 2002; Tsuruoka et al, 1995; Wang and Shearer, 2015; Thomas et al, 2012; Vidale et al, 1998; Bucholc and Steacy, 2016; Tanaka et al, 2002; Metivier et al., 2009). Nevertheless, as there is also no consensus on the possible mechanism for tidal-stress triggering of earthquakes, it is premature to exclude the influence of celestial bodies other than the moon and sun without first conducting a statistical analysis like that presented in this paper.

Only one study has ever attempted to use rigorous statistical methods to evaluate the hypothesis that the conjunctions of the planets might influence earthquake activity. Romanet (2023) evaluated the activity of global earthquakes with magnitudes greater than 7 over 69 years (1950-2018) with respect to the time that planets spent within three degrees of syzygy, or conjunction, with any other planet. Romanet (2023) was not able to reject the hypothesis that large (M7+) earthquakes follow a binomial law during the time period tested, making it extremely unlikely that earthquakes are linked with planetary conjunctions of any type.

Statistical Parameters

This work uses a chi-squared test of independence to determine if days that include earthquakes greater than or equal to magnitude 7 (termed M7+ in this paper) from 1960-2024 are truly independent of the position of Earth in its orbit around the sun. In contrast to Heaton's work and almost all peer-reviewed scientific publications with the exception of Romanet (2023), this study makes no attempt to model and/or predict the outcome of tidal stresses involved in triggering earthquakes. Instead, this work is a statistical study of the timing of earthquakes with relation to the positions of the sun, the moon, and other celestial bodies. Also, like Romanet (2023) and in contrast to other work, this study includes bodies other than the sun and the moon: Jupiter is addressed first; Venus, Saturn, and Mars thereafter.

This work both expands upon and narrows the focus of Romanet's (2023) statistical analysis by restricting the evaluation to planetary orientations involving Earth and by expanding the range of time periods tested to include intervals throughout Earth's entire synodic periods rather than just the conjunctions themselves. This chi-squared analysis uses two different characters of Earth-days: 1) days with or without M7+ earthquakes; and 2) days inside or outside an interval of time defined by Earth's position relative to Jupiter, Venus, and Saturn, respectively. To this end, this study breaks up Earth's orbit into days offset on either side of a reference Earth-Sun-Planet orientation, or zero-point: In one set of chi-squared calculations, the inferior conjunction of the Earth with the planet and the sun is used as a zero-point (termed ONSIDE in this work); in the other set of chi-squared calculations, opposition, or the day when Earth passes behind the sun directly opposite the planet is used (termed OFFSIDE in this work).

For example, if one wanted to determine if M7+ earthquake activity was occurring preferentially during the inferior conjunction with Jupiter, one could use a chi-squared test to count M7+ earthquakes that occurred during any specific interval of time that includes the inferior conjunction (e.g., a 30-day interval starting 5 days before inferior conjunction, "ONJUP -5") over a length of time that includes many synodic periods (e.g., 1960-2024); using that same length of time (1960-2024), one would also count the number of M7+ earthquakes that occurred outside that interval. These counts, which are shown in table 1, are the inputs for the chi-squared test of independence.

| EXAMPLE INTERVAL (30-day interval ONJUP -5) | Days inside the interval | Days outside the interval |
|---|---|---|
| Days w/ M7+ earthquake | 43 | 796 |
| Days w/o M7+ earthquake | 1756 | 21146 |

Table 1: Example inputs for a chi-squared analysis of the 30-day interval starting on ONJUP -5 (5 days before inferior conjunction) for the period 1960-2024.

The chi-squared test of independence, like many other statistical tests, mathematically describes the likelihood that a given set of data violates a null hypothesis. In this case, the null hypothesis is that these categories of days are independent; that is, days with M7+ earthquakes have no relationship with the interval chosen and hence with Earth's position relative to Jupiter. As with other statistical tests, the test-statistic is used to determine a p-value or probability of independence. For a p-value of 0.05, which is the typical value chosen to describe significance, the chi-squared value is 3.841. Because of the controversial nature of this work, this paper utilizes the more strict p-value of 0.02 and chi-squared of 5.412 to determine independence. The p-value for the specified interval in the example, which consists of 30 days starting at Offset Day -5 relative to ONSIDE Jupiter, is 0.006, well below the p>0.02 cutoff. This chi-squared calculation indicates that the earthquake days in this particular interval violate the null hypothesis; that is, days that include M7+ earthquakes are *not independent* but are instead related to Jupiter's position in Earth's sky. Interestingly, M7+ earthquake activity in this interval is not elevated, but is instead *significantly lower* than the average during 1960-2024.

The result described above does not contradict the finding of Romanet (2023), because in the case of Jupiter, the interval of 30-days extends to an angle of approximately 30 degrees out from inferior conjunction, which is much wider than Romanet's (2023) 3-degree analysis. Rather this work corroborates the conclusion of Romanet (2023) that M7+ earthquakes are not more likely to occur during Earth's conjunction with Jupiter.

One example of chi-squared dependence is not enough on its own to show a relationship between the timing of M7+ earthquakes on Earth and the position of Jupiter. One example alone may merely be a mathematical artefact, a product of data clumping or some other unknown statistical foible. Therefore, this work utilizes a computer program to sample earthquake and ephemeris data for the last 64 years (1960-2024) with the purpose of conducting a chi-squared test-of-independence for all intervals (5 to 45 days) spanning the entirety of Earth's synodic period relative to Jupiter, Venus, Mars, and Saturn. This analysis was done to observe any patterns in significant intervals identified by the chi-squared analysis. Weak, chaotic, or sparse patterns of significance would successfully rule out the possibility that the orientation of these planets could be triggering M7+ earthquakes.

Methods:

The computer program (https://github.com/ericonice/the-crazy-idea) written to complete the chi-squared analyses reported in this paper queries two data sets: 1) NASA Horizons ephemeris data (https://ssd.jpl.nasa.gov/horizons/) to determine the dates of inferior conjunction and opposition for Venus, Mars, Jupiter, Saturn, and Earth's moon (Luna) between January 1, 1960 and December 31, 2024; and 2) the USGS Earthquake catalog of M7+ earthquakes from 1960 to 2024 (https://github.com/usgs/devcorner). The program calculates "offset days" referencing either inferior conjunction (ONSIDE) or opposition (OFFSIDE) as the zero point and then determines which of those offset days had M7+ earthquakes and which did not. It then breaks out intervals of 5 to 45 days starting on every offset day from -50 to +300 for both ONSIDE and OFFSIDE Jupiter and then calculates the chi-squared value for that interval (Figure 1, Table 2). The program accounts for days in 1960 prior to the first reference day by determining their offset position relative to reference days (inferior conjunction and opposition) in 1959 and then including them fully into the analysis. Thus all days in the interval 1/1/1960 to 12/31/2024 are included *and no others*.

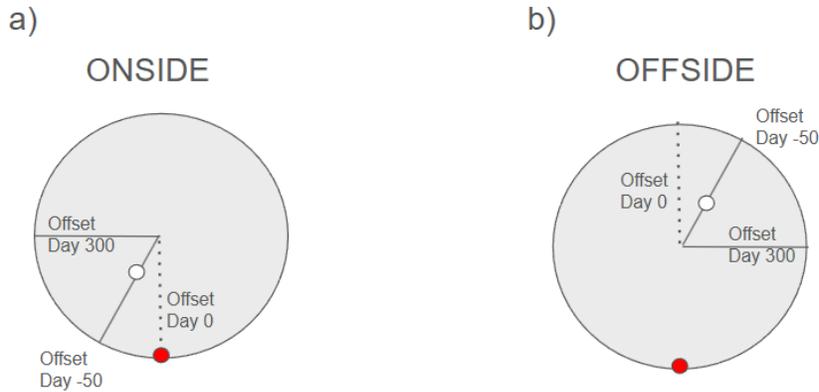

Figure 1. The diagram at left (not to scale) schematically displays the positions of Earth (open circle), Jupiter (red circle), and the sun (center of the grey circle). Offset Day 0 (dashed line) represents the inferior conjunction; Offset Days -50 to +300 (solid lines) show the start-days of intervals considered in the chi-squared analysis for ONSIDE Jupiter. Those intervals range in length from 5 to 45 days. As the synodic period for Jupiter is 399 days (the time between inferior conjunctions), this selected range of offset days avoids overlap. As a consequence of this selection, the part of Earth's orbit between day +300 and day +399 is under-represented in the statistical analysis. In order to account for this and to allow a valid statistical comparison of ALL intervals, this chi-squared analysis is reproduced for the same sets of days relative to OFFSIDE Jupiter (right) in which opposition represents the zero point. In the OFFSIDE chi-squared analysis, the intervals between ONSIDE Offset Day (ONJUP) +300 and +399 are fully represented. These offset days roughly correspond to OFFJUP +100 to +150.

|  | OFFJUP +59 | ONJUP -35 | ONJUP -5 | ONJUP +59 | ONJUP +141 |
|---|---|---|---|---|---|
| $O_{Ai}$ | 79 | 79 | 43 | 63 | 68 |
| $O_{Bi}$ | 1721 | 1721 | 1756 | 1707 | 1702 |
| $O_{Aii}$ | 760 | 760 | 796 | 776 | 771 |
| $O_{Bii}$ | 21181 | 21181 | 21146 | 21195 | 21200 |
| $E_{Ai}$ | 63.6115 | 63.6115 | 63.5761 | 62.5513 | 62.5513 |
| $E_{Bi}$ | 1736.3885 | 1736.3885 | 1735.4239 | 1707.4487 | 1707.4487 |
| $E_{Aii}$ | 775.3885 | 775.3885 | 775.4239 | 776.4487 | 776.4487 |
| $E_{Bii}$ | 21165.6115 | 21165.6115 | 21133.5761 | 21194.5513 | 21194.5513 |
| $\chi^2$ | 4.176 | 4.176 | 7.469 | 0.004 | 0.532 |
| p | 0.041 | 0.041 | .006 | 0.952 | 0.466 |

Table 2. Inputs and results of chi-squared calculations for five selected Jupiter intervals discussed in this paper. Abbreviations are as follows: OFFJUP-Offside Jupiter; ONJUP-Onside Jupiter; E-expected outcome; O-observed outcome; category i-"day inside named interval"; category ii-"day outside named interval; category A-"day with at least one M7+ earthquake; category B-"day without any M7+ earthquakes. The offset day (ONJUP or OFFJUP) indicated above each column is the beginning of a 30-day interval. Also included in these calculations were the following values for the period between 1/1/1960 and 12/31/2024: 839 total M7+ days; 23741 total days with or without M7+ earthquakes.

Comprehensive data tables for ONSIDE and OFFSIDE Jupiter, Venus, Saturn, and Mars chi-squared inputs and results may be found on the Harvard dataverse (Holt, 2025: https://doi.org/10.7910/DVN/SIQXIU). Inputs and results for five selected intervals relative to Jupiter and discussed in this paper are shown in Table 2. Over the period January 1, 1960 to December 31, 2024, there were 23,741 days and 885 M7+ earthquakes; the ONSIDE and OFFSIDE offset days relative to Jupiter that are associated with these 885 M7+ earthquakes, as well as the Sun-Observer-Target (SOT) and Sun-Target-Observer (STO) angles for Venus, Mars, Jupiter, Saturn, and Luna are located on the Harvard dataverse (Holt, 2025: https://doi.org/10.7910/DVN/OH53RL). A complete listing of the calendar dates for Offset Days relative to ONSIDE and OFFSIDE Jupiter, Venus, Saturn, and Mars between January 1, 1960 to December 31, 2024 are included on the Harvard dataverse (Holt, 2025: https://doi.org/10.7910/DVN/57RVDB).

Conducting the chi-squared analysis described above twice, once for OFFSIDE Jupiter (opposition) and once for ONSIDE Jupiter (inferior conjunction) is important, because there are edge effects associated with the most negative and the most positive days that cause undercounting of the significant intervals; some days in the beginning and the end of the analysis are not sampled by every interval length (5 to 45 days). Doing both sides helps to visualize any significance created by this bias, because the undersampled intervals for ONSIDE are fully sampled by the OFFSIDE analysis and vice versa (Figure 1). Also, the synodic period of Earth with respect to Jupiter varies somewhat from year-to-year (395 to 403 days over the period 1960-2024), making the relationship between offset day and Sun-Earth-planet angle less certain for offset days that are farther from the reference zero point. Thus ONJUP and OFFJUP days may differ slightly from a simple, additive relationship (Holt, 2025: https://doi.org/10.7910/DVN/57RVDB); in each case, the closer reference should be used if the goal is to more accurately represent the Sun-Earth-planet angle for these earthquakes that are pulled from different years.

One might argue that it would be better to use the angles between the planetary bodies (Sun-Observer-Target or SOT-angles) to conduct the chi-squared analysis rather than Earth days, and this study does discuss the possible importance of these relative angles in generating patterns of chi-squared significance. However, using SOT- or STO-angles to conduct the chi-squared analysis itself would not result in a valid chi-squared test of independence. The chi-squared test only works to test the independence of different categories of the same subject: angles and days are not the same subject and therefore cannot be tested directly using the chi-squared analysis. The same argument explains why we did not use earthquakes and days; earthquake-days and other days are both different categories of days, but earthquakes and days are not the same subject. Once certain days are recognized as significant by the chi-squared analysis, then any patterns of angles or earthquakes relative to those days are also identified as possibly significant. Therefore even if this statistical analysis cannot treat angles and earthquakes directly, it can still give valuable information about them by identifying the significant days associated with particular angles or earthquakes.

Another possible source of contention is that this study does not include any data before 1960 in the chi-squared analysis. This was done purposefully in order to preserve a pool of information with which to test any hypotheses generated by the patterns of significance detected by chi-squared analysis. Post-1960 M7+ earthquake data worldwide is a well-curated and comprehensive data set (Michael, 2014), making any statistical observations about it more likely to be valid. Seismic data from before 1960 will be addressed once this work evaluating the chi-squared analysis itself is complete on its own to the extent that credible hypotheses can be made about what patterns of M7+ activity are expected during those years given these findings for the years 1960-2024.

There are numerous reasons that this study only utilizes the data from earthquakes magnitude 7 and above. After all, one might argue that if M7+ earthquakes are being triggered by tidal mechanisms, then all other magnitudes should also be triggered, and therefore it is important to explain the decision to stick to larger earthquakes. This is a partial list of the reasons to limit this initial statistical analysis to M7+ earthquakes: 1) The ISC-GEM catalogue as represented by the USGS database is likely to be complete for M7+ by 1960 (Michael, 2014); 2) The number of days with a M7+ is low enough (839 from 1960-2024) that it is possible to debug the computer program by computing the chi-squared analyses by hand for many intervals; 3) This method preserves the smaller earthquakes as another, independent body of data with which to test any possible tidal correlations identified for larger earthquakes, much as the data before 1960; 4) limiting the analysis to M7+ is one method of declustering the data by eliminating most foreshocks and aftershocks from the dataset (Tsuruoka et al., 1995); Finally, as the mechanisms by which tidal forces might trigger M7+ earthquakes are not yet well understood (Beeler and Lochner, 2003; Houston, 2015; Tanaka et al., 2004; Brinkman et al., 2014; Bucholc and Steacy, 2016), earthquakes of lower magnitudes may or may not be mechanistically analogous, which could impact statistical analyses of tidal triggering. With respect to this last assertion, several authors over the past few decades have observed tidal triggering of earthquakes only within specific ranges of magnitudes, either large or small depending on the study (Tanaka et al, 2002; Ide et al., 2016; Lin et al., 2003).

RESULTS

The first result of this work is to show that statistically significant intervals of non-random M7+ earthquake activity are concentrated into two distinct time periods defined by Earth's position relative to Jupiter and the sun (Figure 2). The intervals are not scattered about, as they might be if this were some unfortunate coincidence or an artefact of the sampling method. Instead, the significant intervals occur in two narrowly defined peaks over two particular arrangements: the inferior conjunction and the "preceding neap" position (see diagrams Fig. 2). There is no peak associated with opposition; nor is there a peak associated with the "following neap" position, which is analogous to the preceding neap position but occurs as Earth orbits away from inferior conjunction rather than towards it. Almost all of the significant intervals associated with the preceding neap position occur immediately before Earth reaches the maximum distance from the line connecting Jupiter and the sun. In contrast, the significant intervals that straddle the inferior conjunction occur primarily *after* the conjunction occurs.

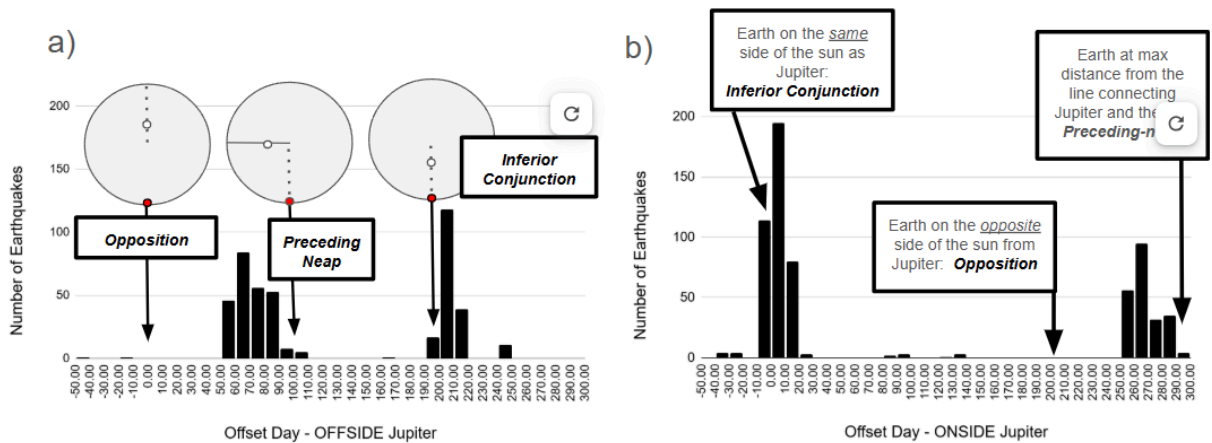

Figure 2. Histograms of the number of intervals (5 to 45 days) returned by the program as significant (p<0.02) for M7+ activity dependent on the position of Jupiter for OFFSIDE (a) and ONSIDE (b) Jupiter from 1960-2024. Intervals are counted by their Offset start-day; for example, the nearly 200 intervals reported between Offset Day 0 and 10 for ONSIDE Jupiter (ONJUP 0 to +10) represent all of the intervals between 5 and 45 days long that started on any day from ONJUP 0 to ONJUP +9. Note these peaks do not necessarily indicate elevated M7+ activity, because some of the intervals identified as significant have unusually low numbers of M7+ earthquakes. The significant intervals are not exact duplicates on the ONSIDE and OFFSIDE graphs, because each excludes start days +301 to +399 to avoid interval overlap (see explanation for Fig. 1).

The next primary result is to show that the length of interval with the greatest significance according to the chi-squared analysis is 30 days (Figure 3a), which is highly suggestive that intervals including a full lunar cycle are more likely to be found significant by the chi-squared analysis. A histogram of interval lengths significant at p≤0.02 for ONSIDE Jupiter illustrates that these data are well organized about an average of 28 days, with a slight skew towards values defined by the two modes at 29 and 34 days in length. The steady increase in the number of significant intervals on either side of the peak of significance indicates that the best measures of center would either be the first mode of 29 days or the average of the two modes: 32 days. Thus, the best centers of this distribution correspond to the length of the lunar cycle. The data for OFFSIDE Jupiter significant intervals corroborate the findings for ONSIDE Jupiter with an average of 27 days and two modes at 29 and 32 days. It does not seem possible that it should be a coincidence that the most numerous interval length returned by the chi-squared statistical analysis should be nearly identical to the lunar cycle. Instead, it is more likely that moon phase plays some role, despite the observation that there is no correlation of global M7+ activity and moon phase when this data is evaluated as a whole unit. Perhaps the tidal effect of Jupiter is most readily found significant by the chi-squared analysis when the effects of the moon are averaged over a lunar cycle. If this is the case, then the implication is that moon phase is significant, but contributing to M7+ activity in a way that cancels out over a full synodic period.

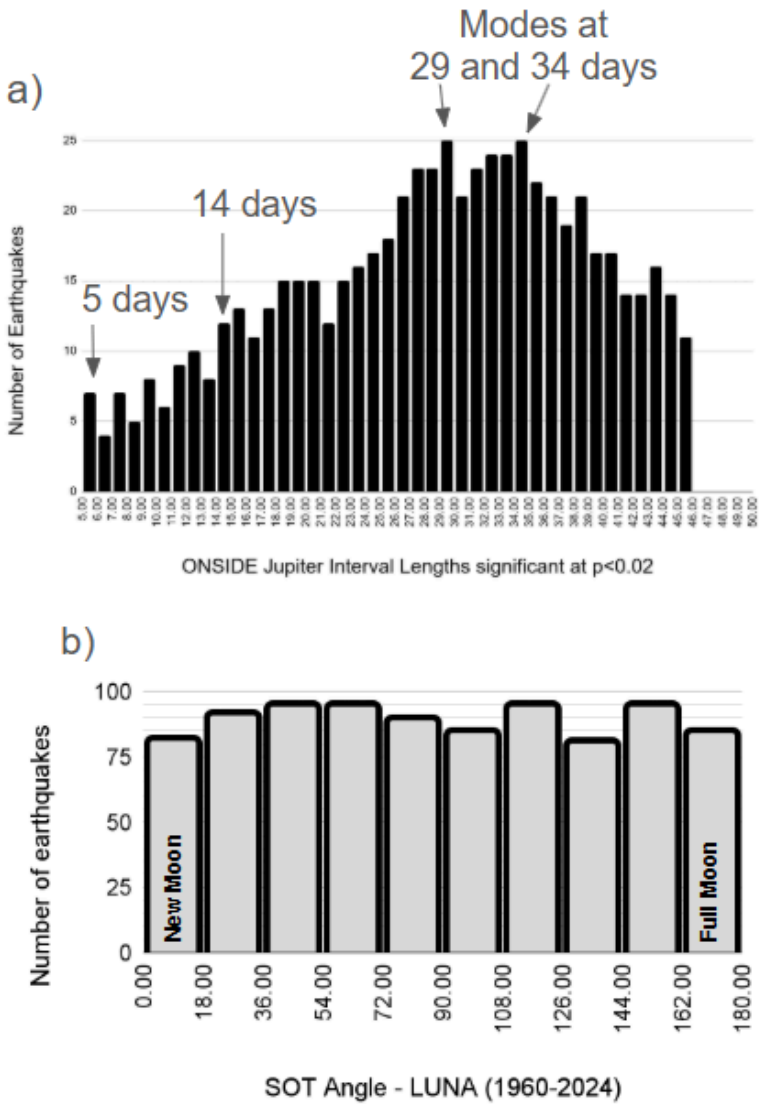

Figure 3. (a) Histogram of the lengths of intervals identified as significant (p<0.02) for ONSIDE Jupiter. The average length of significant intervals is 28 days, and the two modes are at 29 and 34 days respectively. The number of intervals identified as significant steadily increases with interval length from 5 days to the mode at 29 days, where it flattens out until the second mode at 34 days, whereafter it steadily decreases at roughly the same rate as it previously climbed. The center of these data is best represented by either the first mode at 29 days or the average of the two modes at 31.5 days. Thus, every appropriate measure of center for these data corresponds very closely to the lunar cycle. (b) Histogram of the Sun-Earth-Moon (SOT) angles on the days for all 885 M7+ earthquakes from 1960-2024. Full moon and new moon phases are shown for reference. Note that on this histogram, both first quarter and third quarter moons are represented at angles between 72-108 degrees.

In order to visualize how the patterns of significant intervals are expressed against the backdrop of non-significant intervals, the M7+ frequency as a percent of average [M7+ frequency = (number of earthquakes in an interval/total days in that interval)*100/(885 earthquakes/23741 total days 1960-2024)] is plotted against Offside Jupiter Offset Days (Fig. 4). The conventional p≤0.05 is used to distinguish significant intervals (red "x"s), from non-significant intervals (p>0.05; black dots). The continuous and organized smoothing of the data into well-defined and significant peaks and valleys that is brought about by steadily increasing the length of the interval to 30 days is likely a result of averaging out the effects of the lunar cycle (29.5 days). The 30-day interval is more organized and identifies clearly defined highs and lows of M7+ activity for both ONSIDE and OFFSIDE reference points (Figure 5).

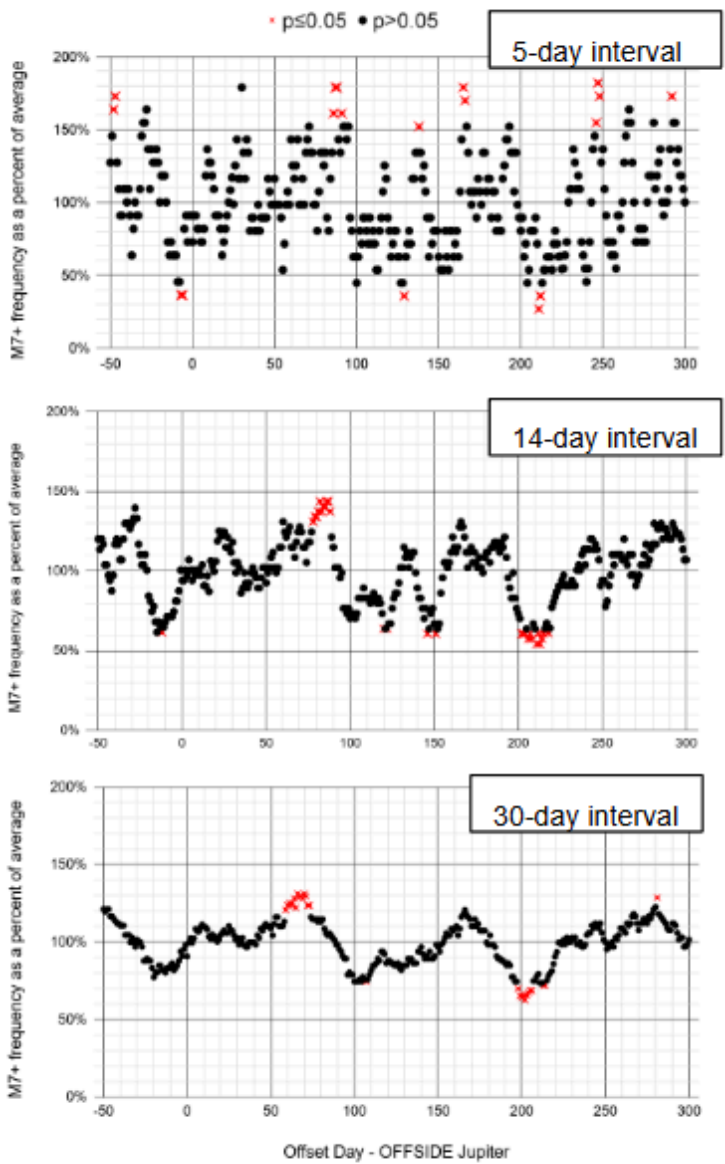

Figure 4. Interval earthquake frequency as a percentage of the average (see text) is plotted against Offset day relative to OFFSIDE Jupiter for all 5-day (top), 14-day (middle), and 30-day (bottom) intervals. Intervals identified as significant to p≤0.05 are shown by the red "x"s and non-significant intervals (p>0.05) are shown by the black dots. Because the offset day represents the start of the interval, the value associated with each point is extended *to the right* of each data point for 5, 14, and 30 days, respectively.

ONSIDE and OFFSIDE representations of the chi-squared data displayed on Figure 5 generally display analogous patterns of peaks and valleys, as they should for they contain largely the same data starting from different reference points. One notable difference is the cluster of 30-day intervals with elevated M7+ activity that immediately precedes the inferior conjunction; these intervals are shown as significant on the ONSIDE graph (FIg. 5b) but not the OFFSIDE graph (Fig. 5a). Figure 5 contains no edge effects associated with the undercounting of significant intervals (see Fig. 2). Differences between OFFSIDE and ONSIDE representations in figures 5a and 5b result only from the variability of Jupiter's synodic period; therefore, the closer ONJUP reference (Fig. 5b) more accurately represents the significance of the clustered intervals preceding the inferior conjunction.

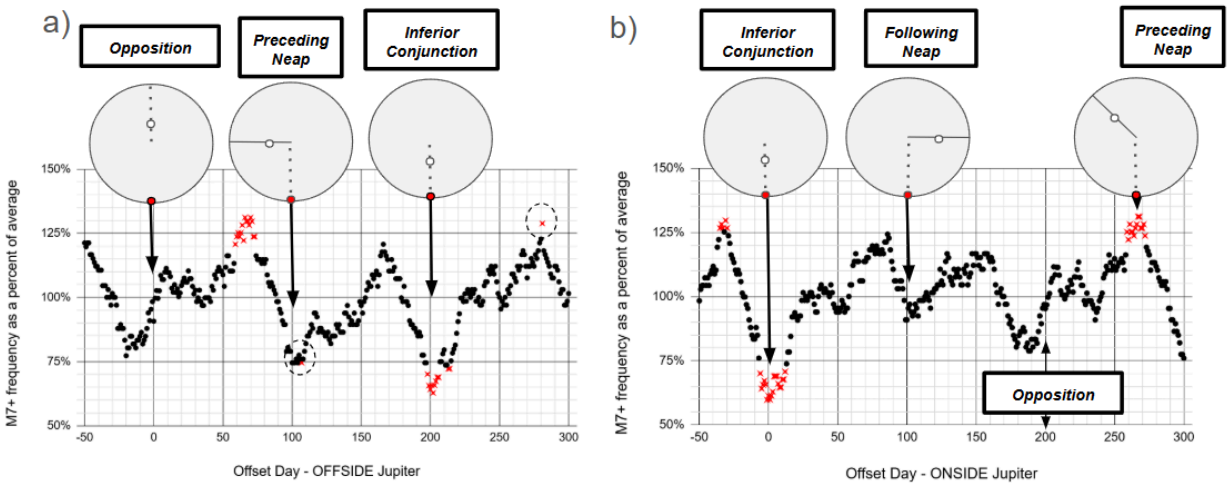

Figure 5. Earthquake frequency as a percentage of the average for a) OFFSIDE and b) ONSIDE 30-day intervals that are included in the chi-squared analyses. Schematic illustrations of the orbital positions of Earth, Jupiter and the sun are as in Fig. 1; red "x"s and black circles are as in Fig. 4. Labels denote the positions of: opposition, inferior conjunction, and both preceding and following neap. Dashed circles denote isolated significant points discussed in the text. Comparing these diagrams, which have the same horizontal scale shown in Fig. 2, illustrates that the two most significant periods are the inferior conjunction and the preceding-neap interval.

Taking ONSIDE and OFFSIDE representations in Fig. 5 together, the chi-squared analysis strongly identifies two significant periods of elevated M7+ activity at the preceding-neap position and inferior conjunction; the data show that M7+ activity first increases (>125% of average) and then sharply decreases in a pulse-like fashion, with those lulls in M7+ activity (<75% of average) lasting about a month. The strongest evidence for significant *elevation* of M7+ activity starts immediately before the preceding-neap position (44 days from OFFJUP +59 to +102), and the strongest evidence for depression of M7+ activity is at the inferior conjunction (49 days from ONJUP -6 to +42). The preceding neap and inferior conjunction positions are the only ones represented by clusters of significant intervals at all three interval lengths: 5-day, 14-day, and 30-days (Fig. 4). Whereas lulls in M7+ activity located at the following-neap position (ONJUP +100) and at opposition (OFFJUP 0) are also present for all three interval lengths, they are not significant at p≤0.02. Also, the lull near opposition is not analogous to that at inferior conjunction, because it does not straddle OFFJUP 0 but instead precedes it by 15 days.

DISCUSSION

Ranking Significant Intervals
Observing the frequency plots for OFFSIDE and ONSIDE Jupiter side-by-side (Fig. 5) allows us to rank the data for significance by including the range of significance as part of the criteria. For example, the isolated significant interval at OFFJUP +281 (dashed circle above 100%) is likely to be an anomalous result: it is not part of a cluster of significant intervals; it is not represented as significant for both ONSIDE and OFFSIDE representations; and it is very far (+281) from the zero point reference of opposition. In contrast, the isolated significant interval at OFFJUP +107 (dashed circle below 100%) is relatively close to the zero point reference of opposition and is part of a broader depression in M7+ activity, which includes at least a dozen offset days that start intervals displaying M7+ frequency less than 70% of average. Consequently, this depression of M7+ activity is considered to be a valid and significant response to the elevation in M7+ activity that occurs just prior (Fig. 5a). Finally, as mentioned previously, the elevated M7+ activity prior to the inferior conjunction is considered to be validly significant even though it is not represented as such on the OFFSIDE plot (Fig. 5a); there is a large cluster of similarly elevated M7+ activity identified as significant on the ONSIDE plot (FIg. 5b), and the inferior conjunction is the closer, more accurate reference point.

Although the chi-squared analysis is an excellent tool to mathematically describe significance in categorical data, it isn't really the right tool to rank significant peaks and valleys relative to one another. Because the chi-squared test pits everything inside the interval against everything outside, it has trouble identifying any but the largest peaks and the lowest valleys on the graphs like those shown in Fig. 5. It is obvious that there is a predictable cyclicity to these data, with at least one other lull in M7+ activity at ONJUP +187 (immediately before opposition) and an associated peak just prior at ONJUP +160 that together represent the "following-neap position". This paper will argue that this peak and associated lull are mechanistically analogous to the preceding-neap position. However, the chi-squared analysis has no regard for such cycles, and as a result of this, it is unlikely that more of the wave-like cyclical patterns in the above diagrams will be verified as significant, even as more data accumulate with time. Another mathematical tool, such as a fourier analysis, might be more effective at picking out cyclicity in the data now that it has been shown that the M7+ activity is a function of the Offset day relative to Jupiter.

Elevated M7+ activity associations with the preceding-neap interval
The 30-day intervals located at the preceding-neap position comprise the most obviously meaningful period of heightened M7+ activity identified by the chi-squared analysis. This position displays a large, uninterrupted range of significant intervals starting at OFFJUP +59 that encompass a period of elevated M7+ activity spanning approximately 44 days. This period of elevated M7+ activity ends precipitously at OFFJUP Day +100, after which M7+ activity is reduced (75% of average) for another 35-40 days before returning to more average M7+ activity ahead of inferior conjunction. Elevated M7+ activity in the initial 30 days of this period (OFFJUP +59 to +88) is consistent over time, and not grouped into clumps of multiple earthquakes occurring in the same year (Figure 6). Also, the elevated M7+ activity between OFFJUP +59 and +88 is correlated with the full and new moons (Figure 7). Finally, this cluster of significant intervals begins just as Earth passes through a sun-Earth-Jupiter angle of 45 degrees, the point at which Jupiter is oriented at 45 degrees to Earth's principal stress axes as defined by the combined tidal forces of the sun and moon. Homogeneous solids fail at angles of 30-45 degrees to the principal stress axes (Lockner and Beeler,

2002; Savage et al., 1996; Yu and Wang, 2018), which suggests a possible mechanism by which such a miniscule tidal stress such as that generated by Jupiter could trigger an earthquake: namely, the small, but abrupt, perturbation of Jupiter's tidal influence directly along the plane of maximum tidal shear.

Comparative study of three intervals of varying significance
The following discussion uses comparisons between 30-day intervals at three different positions, two of which were identified as significant by chi-squared analysis and the third of which has been ruled out as significant with regards to M7+ activity: 1) the preceding neap position, 2) the inferior conjunction, and 3) the non-significant interval starting on ONJUP +59.  Although the 30-day interval at ONJUP +59 might seem to be located in an analogous position to OFFJUP +59, Earth is actually orbiting away from Jupiter at that point, and the angles are quite different (117-87 degrees for ONJUP +59 vs. 44-74 degrees for OFFJUP +59). M7+ activity is moderately elevated in two of these intervals: OFFJUP +59 (120% of average) and in ONJUP -35 (127% of average); the third interval (ONJUP +59) displays average M7+ activity (105% of average). The characteristics that are evaluated for these three representative intervals in Figures 6-9 are: 1) M7+ consistency over time or lack thereof (clumping; Fig. 6); and 2) the quality of M7+ correlation with moon phase in each interval (Fig. 7).

*Annual consistency of M7+activity*
Qualitatively, a visual inspection of the graphs for the two 30-day intervals starting at ONJUP -35 and ONJUP +59 (right-side of Fig. 6) would dismiss the data as a product of clumping, which could potentially impact the statistical analysis. The graph on the left (OFFJUP +59), however, is harder to discount on the grounds of clumping. Only one fifth of the years since 1960 have had no earthquakes in this 30-day period beginning at OFFJUP +59, whereas about one third of the years display no earthquakes in the other two intervals (ONJUP -35 and ONJUP +59).

A more quantitative argument addresses the distribution of M7+ earthquakes about the mean. The experimental probability that a M7+ earthquake will occur based on historical data (1960-2024) is one every 26.8 days if there are no triggers involved; this was determined by simply dividing 23741 total days by 885 total earthquakes.  Thus, having one earthquake in this 30-day interval each year is not noteworthy, and the 67-79% of years with an earthquake observed in all three of the intervals shown in Fig. 6 is well within the bounds of random chance (Fig. 6).  However, if M7+ distribution in time is randomly distributed about a mean, we would expect at least the same percentage of years to have zero earthquakes as we see having more than one earthquake. This is indeed what is observed for the non-significant interval ONJUP +59 (33% with zero and 30% with two or more). However, the OFFJUP +59 interval has twice as many years with 2 or more (41%) than there are without any M7+ earthquakes (21%). Another way of putting it is that if these earthquakes were evenly distributed about the 26.8 day mean, there should be at least 25 years with no M7+ earthquake for OFFJUP +59; instead, there are only 13 years with no M7+ earthquake. These data show that there is a consistent and moderately predictable elevation of M7+ earthquake activity in the preceding-neap position. Because the elevated M7+ activity in the OFFJUP +59 to +88 interval is consistent over time and not a product of clumping, statistical tools to identify significance are valid in this interval. Therefore, observed elevation of M7+ earthquakes is indeed correlated with the present of Jupiter at a particular position in Earth's sky.

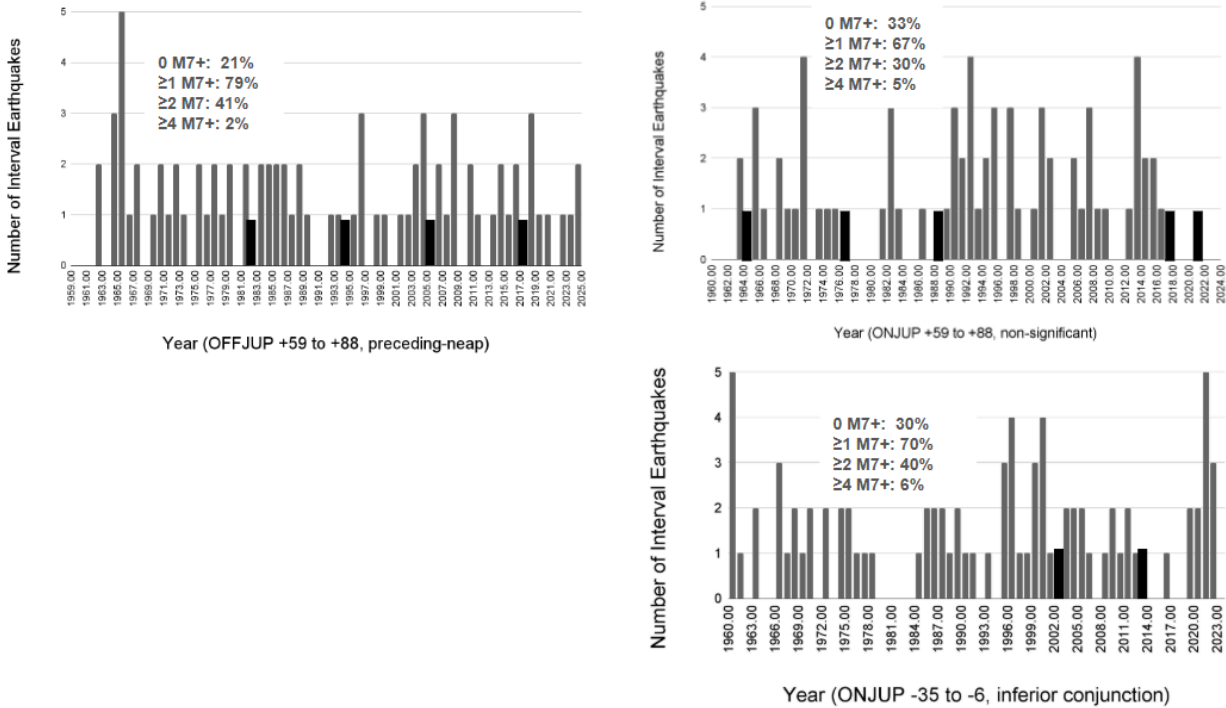

Figure 6. Histograms of the number of M7+ earthquakes that occurred in each of these intervals by year, 1960-2024. The black boxes indicate years that contained less than 15 days of the noted interval and which also did not have any M7+ earthquakes; these years in black are excluded from the total years when calculating the boxed percentages above each plot. These percentages represent the fraction of years 1960-2024 with 0, ≥1, ≥2, and ≥4 M7+ earthquakes in the noted interval.

For ONJUP -35, this elevation in M7+ activity is more clumpy and less obviously dependent upon a particular offset day relative to Jupiter than it is for OFFJUP +59. Although this interval does show 1/3 more years with two or more (40%) M7+ earthquakes than with zero earthquakes (30%), it is not as clearly a departure from what we'd expect from a random distribution (19 observed years of zero earthquakes vs. 25 expected). This is consistent with the results of the chi-squared analysis, which more clearly indicates the inferior conjunction lull as significant when compared to the cluster of significant intervals of elevated M7+ activity that immediately precedes that lull (Fig. 5).

*Evaluation of Correlation with Moon Phase*
Recalling that moon phase is not correlated with M7+ activity when all the data from 1960-2024 are counted (Fig. 3), the variability of M7+ activity with moon phase during these different intervals is noteworthy (Fig. 7). All three intervals show some elevated M7+ activity associated with particular lunar SOT-angles that is outside of one standard deviation from the mean (white vertical rectangles in Fig. 7). In order of degree in this regard: OFFJUP +59 is strongly associated with the full moon and with the crescent phase within 30 degrees of the sun (close to the new moon); ONJUP +59 is significantly associated with the gibbous phase; and ONJUP -35 is only weakly associated with the gibbous phase.  Only OFFJUP +59 and ONJUP +59 intervals show any association with depressed M7+ activity outside of one standard deviation. Both of them show weak associations of depressed M7+ activity at different SOT angles near the quarter moon: OFFJUP +59 at SOT angles of 108-126 degrees (gibbous phase), and ONJUP +59 at SOT angles of 72-90 degrees (crescent phase).

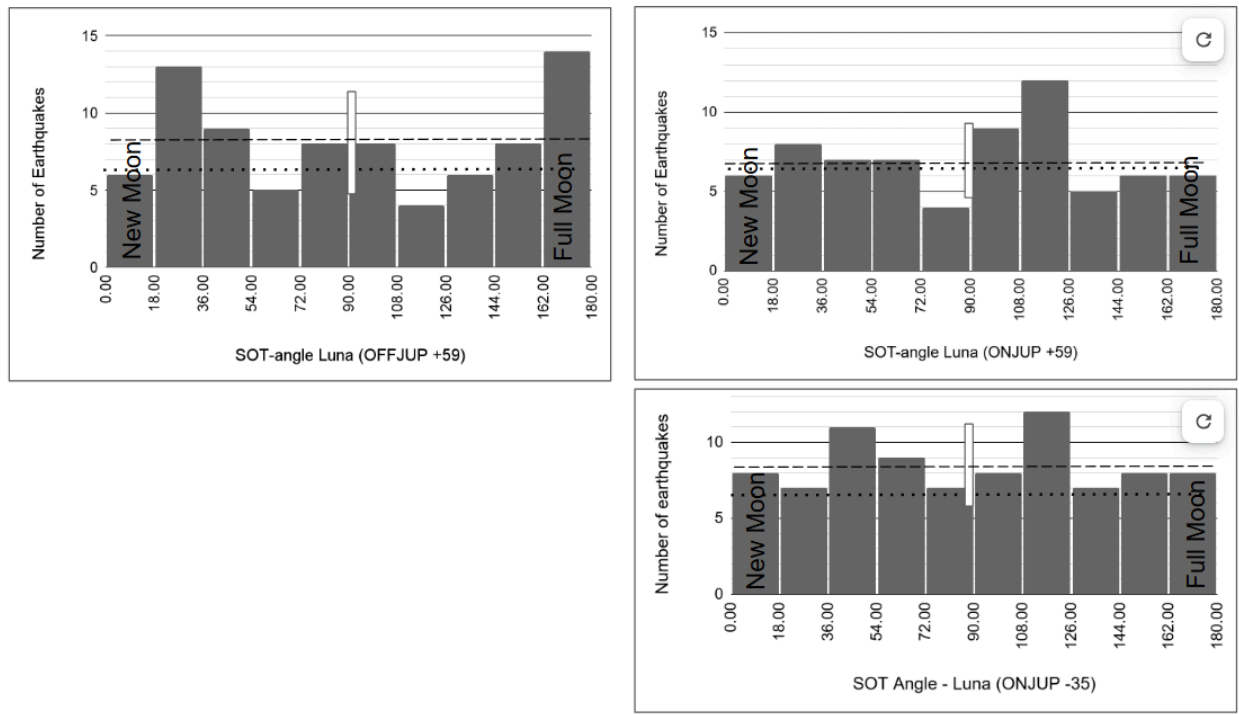

Figure 7. Number of M7+ earthquakes observed in each 30-day interval shown in relation to moon phase, which is represented by the Sun-Earth-Moon (SOT) angle: full and new moon phases are labeled. The dashed lines are the observed number of earthquakes in each interval (81, 69, and 85) divided by 10 to show the number of earthquakes expected in each bin. The dotted line represents what we would expect in each bin if the average number of earthquakes (64) over the period 1960-2024 had occurred in each of these intervals. The white rectangle shows one standard deviation for the number of M7+ earthquakes observed in each bin for each interval.

Notwithstanding the apparent correlations with moon phase for ONJUP -35 and ONJUP +59, binning the data differently for these two intervals results in a chaotic and contradictory pattern of supposed moon phase correlation for each of these intervals. It is likely that the apparent correlation in moon phase for these two intervals, which in figure 9 are binned at 18 degrees, is an artefact of having not enough data to rule it out. However, no matter what bin size is chosen for the OFFJUP +59 interval, the full and new moon phases are consistently more populated than the quarter moon phases, indicating that for this interval, at least, there is a clear correlation of heightened M7+ earthquake activity during the full moon and near the new moon.  The reason this correlation is not visible on the histogram for all 885 earthquakes from 1960-2024 is that the number of earthquakes in the ONJUP +59 interval (81) is less than one tenth that of the entire data set; hence the correlation in this interval is lost in the noise when the data set is reviewed as a whole (see Fig. 3).

The association of increased M7+ activity with full and new moons suggests that it is the *combination* of the position of Jupiter relative to the sun and Earth *along with* the tidal effects of the moon that triggers earthquakes in this interval. Jupiter and the moon in optimal position could occur at any offset day in this 30-day interval; if this is the case, intervals of 30 days that average out the lunar cycle would be more likely to be recognized as significant by the chi-squared analysis, which we do observe (Fig. 3).

Relationship of M7+ activity to Sun-Earth-Jupiter (SOT) angle
Before one can evaluate whether or not earthquakes tend to be correlated with certain Sun-Earth-Jupiter (SOT) angles more than others, it is necessary to consider whether or not the time available for earthquakes to occur is the same at every angle; the short answer to that query, is "no", as the frequency plot in figure 8 plainly shows that Earth spent more time at SOT angles less than 90 degrees and less time at SOT-angles greater than 90 degrees. Thus, linearly relating angles to days over the entire course of the synodic period would be invalid. With regards to this work, OFFJUP +59 to +102 (45-81 degrees) is the longest continuous interval sampled in this study, and the histogram does slope downwards slightly toward the upper part of that range. The slope there is small enough that it doesn't significantly impact any of the results presented herein, as over these short arcs, days may be considered an acceptable proxy for angles. None of these factors influence the chi-squared analysis for any single interval, because that analysis uses days, not angles. However, when comparing the response of M7+ activity to SOT-angles from *different* intervals, one must take these considerations regarding the differences between angle and day into account.

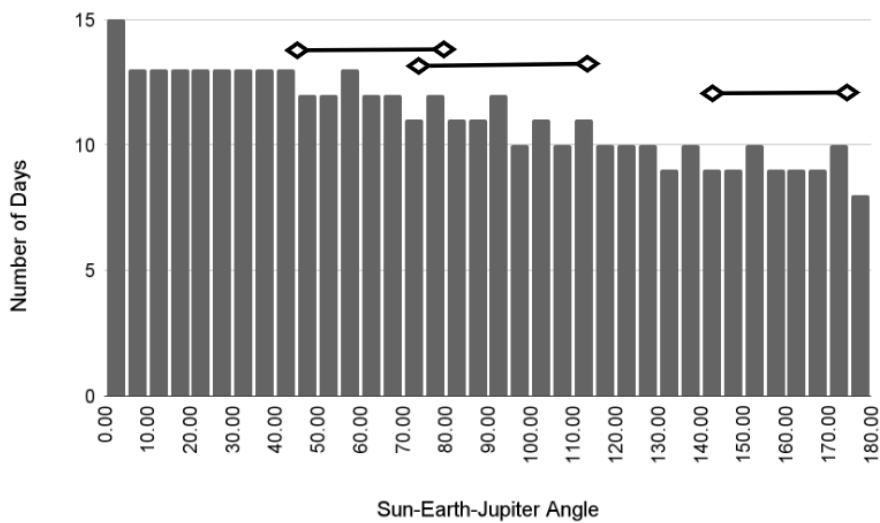

Figure 8. This frequency evaluation was prepared by first determining the SOT angle on every day from 1960-2024; then categorizing the data as to offset day; and finally taking the mean of the angles on each particular offset day. For example, OFFJUP -50 occurred 59 times between 1960-2024; the mean of the SOT angles on those days was 38.32 degrees; this angle along with the angles from a dozen other offset days, makes up the bar (n=13) located between 30 and 40 on this figure. The angles for the intervals discussed in this work are delineated by the black, diamond-tipped lines above the histogram.

Offset days in different years occur at slightly different sun-Earth-Jupiter angles (Fig. 9 and Holt, 2025: https://doi.org/10.7910/DVN/57RVDB), and this variability in angle is in itself not uniform throughout Earth's synodic period.  This variability increases in the months surrounding the inferior conjunction, although error bars contract briefly on the day of the inferior conjunction itself. Offset days are treated equally by the chi-squared analysis. If M7+ earthquakes are triggered at particular angles and some offset days have more of those angles than others, then this has the potential to impact whether a particular interval is singled out as significant by the chi-squared analysis. This is important when evaluating the relative significance of intervals located at the preceding-neap position and the inferior conjunction, as will be seen below. [Note that the variability in SOT-angle with offset day is much less for a graph constructed relative to ONSIDE Jupiter, because Earth's orbit is more consistent relative to Jupiter as Earth is moving away from inferior conjunction towards opposition in the period from 1960-2024; in that case, one-sigma error bars lie inside the thickness of a line drawn connecting the points.]

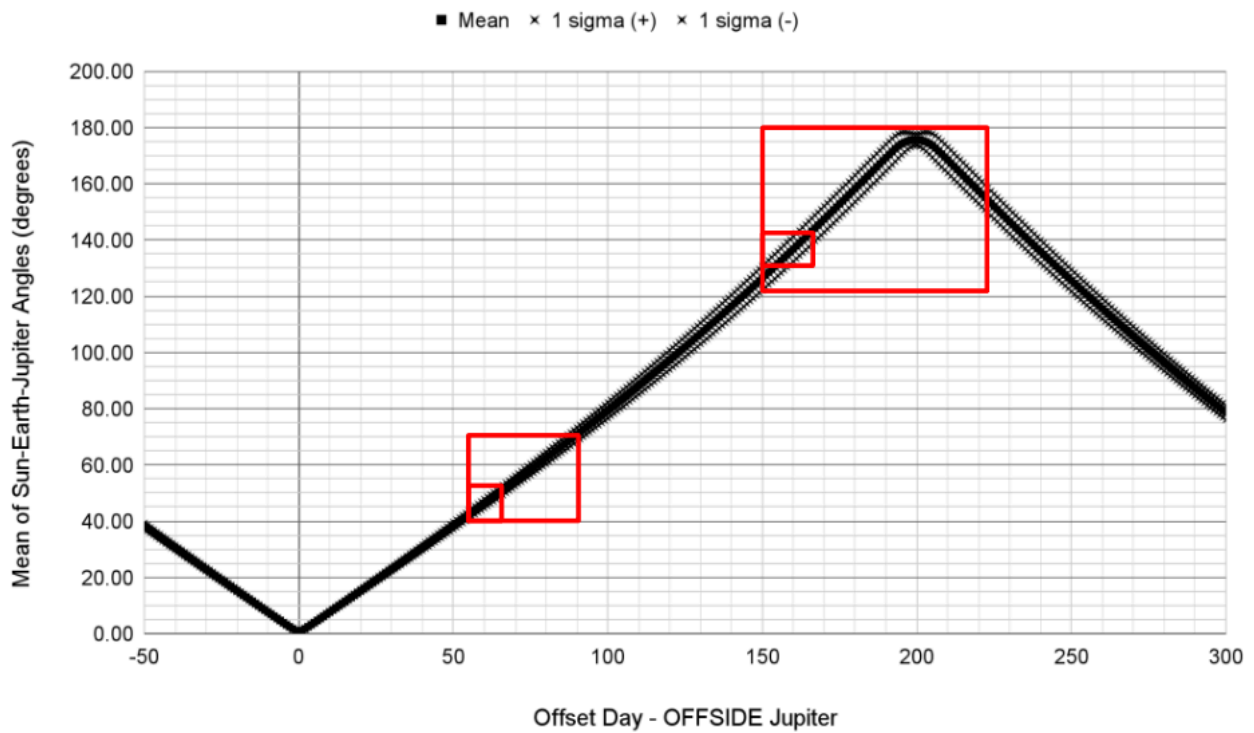

Figure 9. Sun-Earth-Jupiter angles averaged over the period 1960-2024 as a function of offset day relative to OFFSIDE Jupiter. Points above and below the graph at +1 and -1 sigma show the standard deviation of values for angles at each offset day. Four inset diagrams in Figs. 11 and 12 correspond to the areas inside the red boxes.

M7+ activity in the preceding neap position (OFFJUP +59 to +88) and at inferior conjunction (ONJUP -35 to -6) increases as Jupiter's position approaches 45 degrees to Earth's plane of maximum tidal shear due to the sun and  moon (Fig. 10).  In the preceding-neap position, 30-day M7+ activity ramps up above 100% beginning on

OFFJUP +40, where Jupiter is at an SOT angle of 30 degrees (Fig. 9) and peaks as the SOT-angle passes upward through 54 degrees on or about OFFJUP +68 (Fig. 10). Correspondingly, at inferior conjunction, M7+ activity increases steadily through 100% of average on OFFJUP +150 (SOT=126 degrees), which is at an SOT-angle of 54 degrees to the line between the sun and the Earth (Table 3). Steadily increasing M7+ activity prior to the inferior conjunction peaks between 153 and 158 degrees (23-27 degree supplementary angles) between ONJUP -25 and -21, with choppy and variable M7+ activity continuing thereafter (Fig. 10 and Table 3). Thus both the preceding neap interval and the inferior conjunction display M7+ activity that ramps up as Earth passes through an angle of 45 degrees relative to Jupiter and the sun.

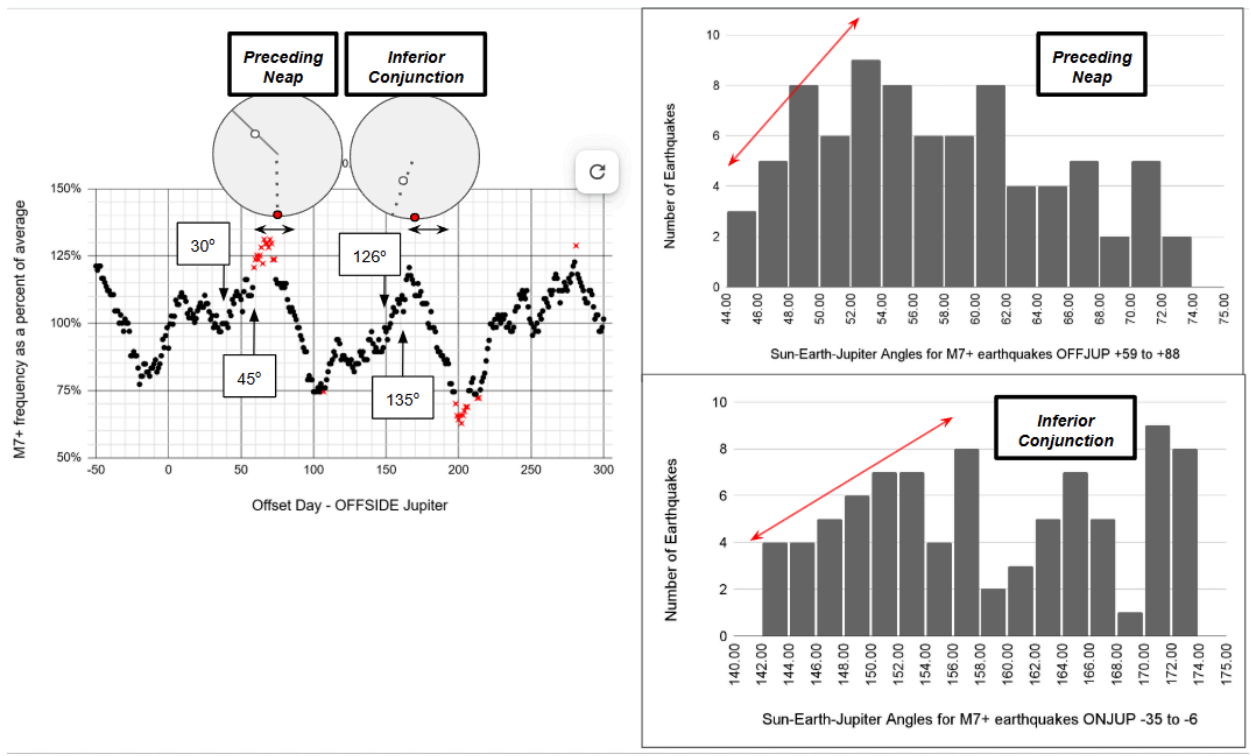

Figure 10. Histograms (right) display the frequency of earthquakes at various Sun-Earth-Jupiter angles during two 30-day intervals of elevated M7+ activity: the preceding-neap position and the inferior conjunction. Frequency plot and orbital diagrams above are as in Figs. 1 and 5 with the dark grey double arrows below the orbital diagrams denoting the length of the intervals displayed at right. The angles shown on the frequency plot show the offset day that Earth passes through that SOT-angle relative to Jupiter. Red arrows above the histograms show the angle-range over which M7+ activity is steadily increasing.

|  | SOT angles at the start of elevated M7+ activity (degrees) | SOT angles at the end of elevated M7+ activity (degrees) | Length of M7+ activity ramp-up | Duration of significant elevated M7+ activity (Fig. 10) |
|---|---|---|---|---|
| Preceding neap | 30 | 49-53 | 25-30 days | 44 days |
| Inferior conjunction | 126 (supplementary: 54) | 153-158 (supplementary: 23-27) | 26-30 days | 35 days |

Table 3: SOT-angles that bound the periods of M7+ activity ramp-up are shown along with the length of that ramp-up in days for the inferior conjunction and the preceding-neap intervals. The total durations of the periods of time identified by the clusters of significant 30-day intervals in the preceding-neap interval and prior to inferior conjunction are shown in the far right column.

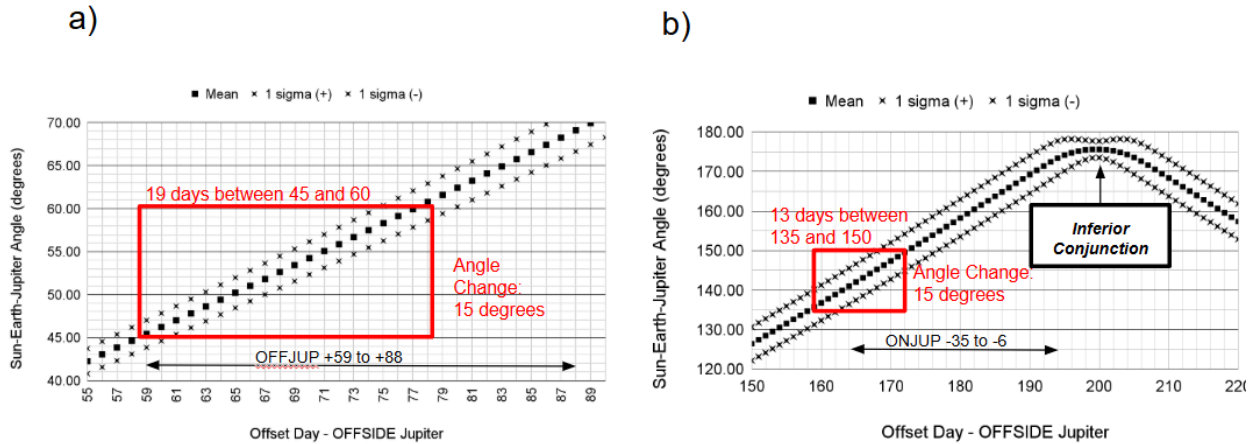

Figure 11. Two inset diagrams from the boxes on Fig. 10 showing the variability in the sun-Earth-Jupiter angle in the vicinity of (a) OFFJUP +59 to +88 and (b) ONJUP -35 to -6. Note these plots are not at the same scale. The red boxes on these plots illustrate the slope of the mean angle relative to offset day with the length of each box displaying the time period over which Earth moves through an angle of 15 degrees.

The path Earth takes relative to Jupiter during its synodic period dictates that the number of days that Earth takes to pass through a given Sun-Jupiter-Earth angle decreases as Earth approaches inferior conjunction: it takes 19 Earth days to orbit through a Sun-Jupiter-Earth angle of 15 degrees in the preceding-neap position and only 13 days to traverse that same arc near the inferior conjunction (Fig. 11). Jupiter is positioned along the plane of maximum shear stress in both of these intervals (STO angles of 45 and 135 degrees), but Jupiter spends 46% more time in that orientation during the neap interval relative to inferior conjunction. Spending a greater amount of time within the critical angles allows the earth time to respond seismically to any perturbation in the tidal force along the plane of maximum shear stress, which could account for the increased total time over which M7+ activity is elevated in the preceding neap interval (44 days compared to 35 days, right column of Table 3) and also for its more consistent elevation in M7+ activity from year to year (Fig. 6). Another possibility is that as a result of the fact that Earth spends a greater amount of time between the critical angles (30-55) in the preceding-neap position relative to that of the inferior conjunction, the chi-squared analysis will rank the preceding-neap position intervals as more significant *even if earthquake activity is equally enhanced relative to the angles in these intervals*.

Further, differences in Earth's synodic period mean that Offset days are more accurately correlated with SOT-angle at particular positions (e.g., preceding-neap vs. inferior conjunction) when ONSIDE reference points are used relative to onside and vice versa. Error bars at one sigma to the mean (Fig. 12) show that the angle variability at any particular offset day in the vicinity of the preceding-neap position (4 days) is less than half that observed near the inferior conjunction (9 days) when the OFFSIDE reference point of opposition is used. Because of these normal variabilities in Earth's synodic period relative to Jupiter, the Earth is consistently located at a Sun-Earth-Jupiter angle of 45 degrees on OFFJUP +59, whereas Earth is less than half as likely to be located at the critical angle of 135 degrees on OFFJUP +157. An analogous representation prepared with the ONSIDE reference shows a much smaller range in the variability of offset days ($\mp 1$ day) at that same position of 135 degrees, indicating that the ONSIDE reference is a more accurate representation of SOT-angle than the OFFSIDE reference for this interval. This difference in the accuracy to which the offset day properly represents a particular SOT-angle largely explains why the inferior conjunction shows significant elevated M7+ activity on the ONSIDE representation but not the OFFSIDE representation; that is, the ONSIDE representation more correctly displays the significance of the SOT-angle in producing elevated M7+ activity.

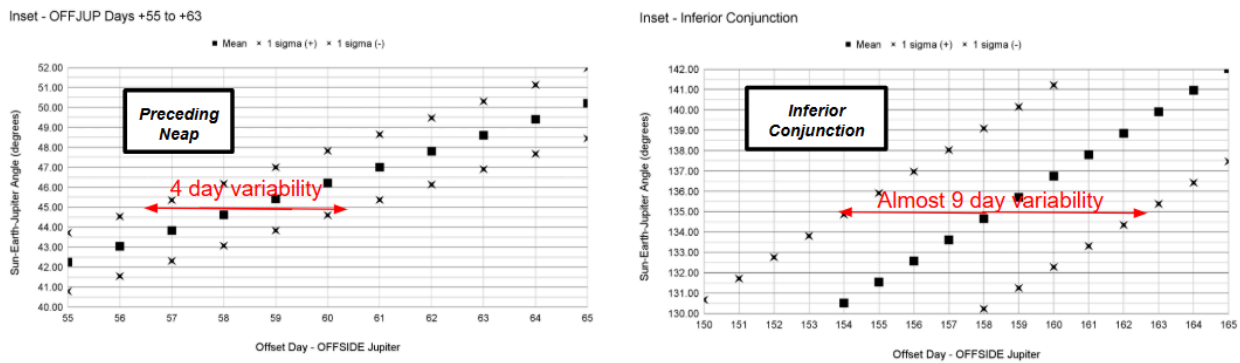

Figure 12. Inset diagrams from Fig. 9 showing the variability in offset day at the critical angles of 45 and 135 degrees for the preceding-neap position and inferior conjunction with OFFSIDE Jupiter as a reference.

The fact that the chi-squared analysis doesn't discriminate between days with regards to how quickly Earth moves through a possible critical angle doesn't invalidate the chi-squared analysis: If earthquakes are triggered at a certain angle, and the Earth spends more time at that angle, then the chi-squared analysis will identify it; if Earth passes through that critical angle so quickly that an insignificant number of earthquakes were triggered, then the chi-squared analysis will very rightly ignore it.  However, this underscores the pitfalls of attempting to use the strength of the chi-squared analysis (number and length of significant intervals) to rank particular periods as more or less significant than one another.  It is possible that the interval at ONJUP -35 to -6 is mechanistically analogous to that of OFFJUP +59 to +88, but the chi-squared analysis will rank intervals near the preceding neap position as more significant for M7+ elevation.

If M7+ earthquakes are triggered by Earth moving through an angle of 45 degrees to the plane of maximum shear stress on Earth as it travels from opposition to inferior conjunction, then that begs the question:  Why are there no intervals identified as significant as Earth moves past inferior conjunction towards opposition?  There are two offset days where this might occur:  1) ONJUP +40 (135 degrees) and 2) ONJUP +141 (45 degrees).  The first is easy to address, and the second will require a more thorough discussion.

Recall that the burst of M7+ activity that immediately precedes inferior conjunction is followed by the most significant period of reduced M7+ activity identified by the chi-squared analysis; this putative relaxation period continues beyond the point at which Earth passes through 135 degrees at ONJUP +40.  The group of 30-day significant intervals with M7+ activity of less than 75% have start-days as late as ONJUP +15 (Figure 5), which means this reduced M7+ activity lasts until at least ONJUP +45, at which point, Earth has already passed through the critical angle of 135 degrees relative to Jupiter.  The close proximity in time between the critical sun-Earth-Jupiter-angles on either side of the inferior conjunction creates an asymmetry in which M7+ activity is elevated before the inferior conjunction and depressed after it.

Symmetry demands a second interval of elevated M7+ activity starting at ONSIDE +141 at a sun-Earth-Jupiter angle of 45 degrees: the following-neap position analogous to the preceding-neap position for Offside Jupiter.  In fact, there is a burst of M7+ activity at that following-neap position. Approximately 17 days of elevated M7+ activity begin at ONJUP +141 at an angle of 47 degrees and continue through ONJUP +158 at an angle of 31 degrees, after which M7+ activity drops sharply into a lull for about a month preceding opposition (Fig. 5).  Although the elevation in M7+ earthquake activity and the lull both display approximately the right amount of time to be analogous to both of the other significant intervals already identified, the effect is not as extreme, and the chi-squared analysis consequently does not recognize it as significant (p=0.466; Table 2).

It could be that the stress reduction during the previous significant intervals means that there isn't enough stress built up to provoke significant M7+ seismic energy release at the following-neap position.  It could also be that there just isn't enough M7+ earthquake data at present for the chi-squared analysis to single out these intervals.  It might be possible to test this by incorporating more earthquake data, either from years before 1960 or from lesser magnitudes, or both in order to see if the chi-squared analysis would then recognize these intervals as significant.

A plot displaying 30-day intervals at p≤0.15 illuminates the similarity in character of the following-neap position to the preceding-neap position and inferior conjunction, but at a lesser degree of confidence (Fig. 13). On figure 13, there are three distinct lulls in M7+ activity, all of which follow the only periods of time in which Earth moves through an angle of 45 degrees relative to the sun and Jupiter and in which M7+ activity is elevated. Along with all of the other data presented in this work, this provides substantial evidence that M7+ earthquakes occur on Earth as a result of tidal perturbations from Jupiter at sun-Earth-Jupiter-angles near the plane of maximum shear stress as generated by tidal forces of the sun and moon.

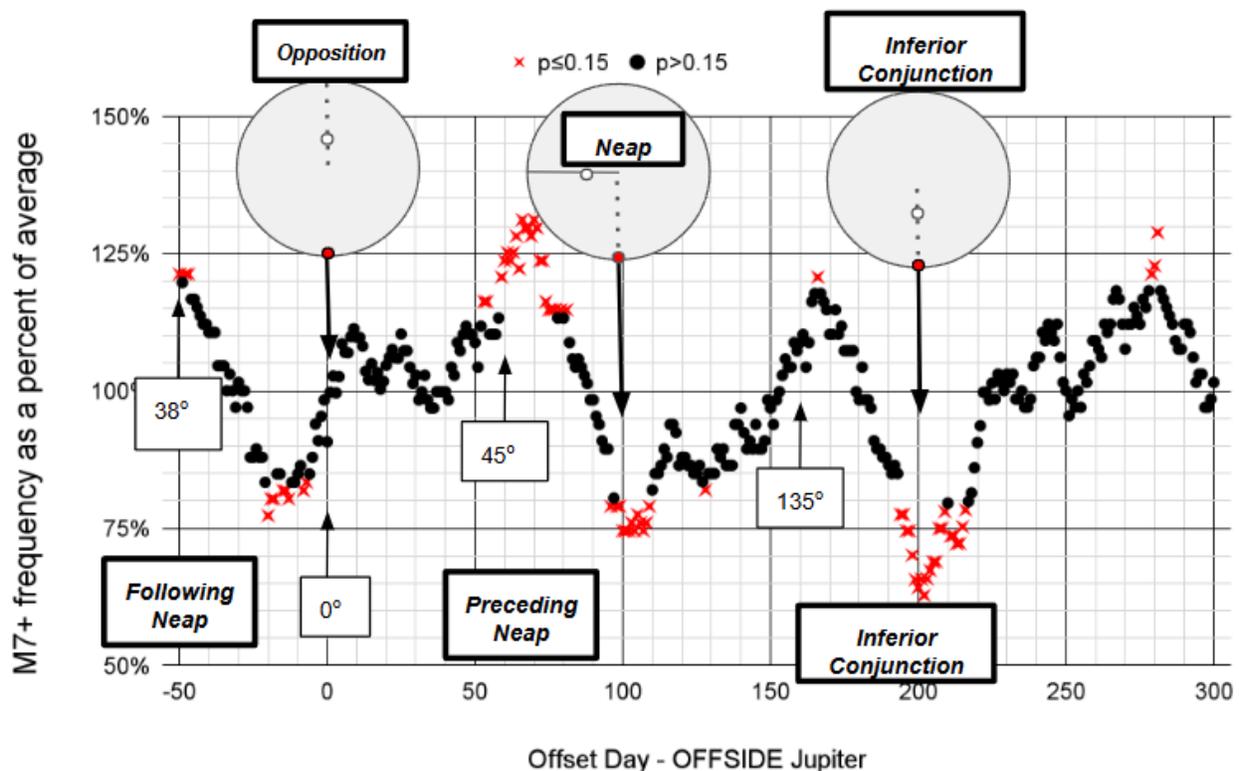

Figure 13. This plot of M7+ frequency for OFFSIDE Jupiter utilizes a lower-confidence p-value (0.15) to illustrate the pattern of three pronounced lulls in M7+ activity referenced to the relative positions of Earth and Jupiter. Schematic diagrams along the top of the plot are as in Fig. 1. An analogous plot for ONSIDE Jupiter cuts off the preceding-neap low and does not include significant elevated M7+ activity prior to the following neap lull. The significant elevated M7+ activity recognized on this plot prior to the following neap lull is probably not significant due to edge effects, analogous to what was observed for the inferior conjunction position on the ONSIDE Jupiter plot (see Fig. 5).

M7+ activity and other planets
In order of decreasing significance, Venus, Saturn, and Mars all contain intervals identified as significant by analogous chi-squared calculations like that conducted for Jupiter (Fig. 14 and Holt, 2025: https://doi.org/10.7910/DVN/SIQXIU). Mars is not included on Fig. 14, because Mars' variable synodic period means that there is not a consistent relationship between offset day and Sun-Earth-Mars-angle for Mars over the period 1960-2024; therefore, even though it does contain a few significant intervals, the error bars are too large on the SOT-angle to draw any conclusions regarding Mars' possible tidal influence on Earth.

There are many analogous features of the frequency plots shown in Fig. 14 for Venus, Jupiter, and Saturn. There is a peak of M7+ activity during the following neap position (SOT angles of 35-55 degrees) for all three of these planets. There is also a pronounced lull in M7+ activity during the inferior conjunction for Venus and Jupiter and to a lesser extent for Saturn.  Just prior to inferior conjunction, there is a peak in M7+ activity for all three planets, although in Venus' case, the lull follows the peak associated with the preceding neap position (SOT-angles of 35-55 degrees) rather than the separate inferior conjunction peak (SOT-angles of 125 to 145 degrees) that is associated with Jupiter and Saturn. The M7+ activity in the preceding neap position that is observed so strongly in association with Venus and Jupiter, is only weakly displayed by Saturn, if at all. The poor correlation for Saturn is undoubtedly related to its weaker tidal influence; it has less mass than Jupiter and is almost twice as far away.

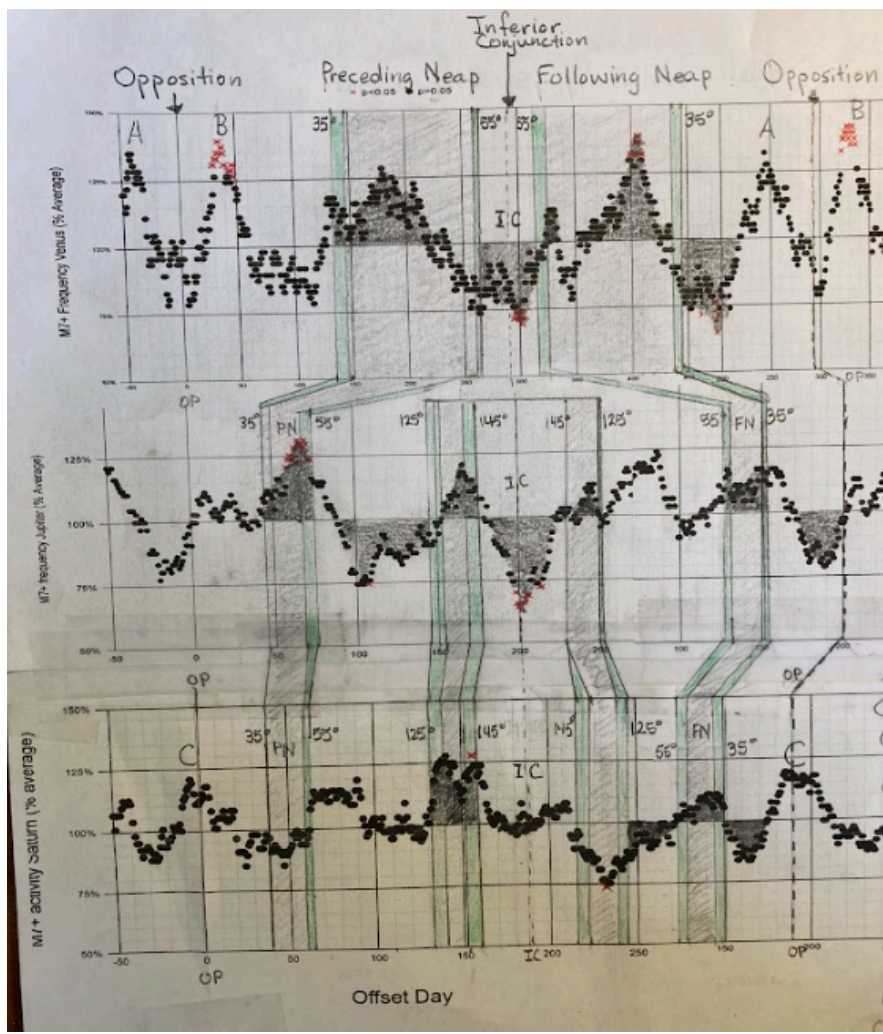

Figure 14. M7+ frequency as a percent of average for, from top to bottom: Venus, Jupiter, and Saturn. Offside and Onside graphs have been cut-and-pasted to show the entire synodic period for each planet. Letters A, B, and C above the graph for Venus and Saturn denote which peaks are repeated as the graph wraps around upon itself. Red and black symbols are as in Figs. 4 and 5. Lines and lightly shaded areas connect SOT-angle ranges of 35-55 degrees and 125-145 degrees between each graph. Dark shaded areas above and below the 100% line highlight elevated M7+ activity within these ranges or depressed M7+ activity outside these ranges. Green shading represents the 1 sigma error bars on the SOT-angle for any given offset day; vertical lines without green have error bars thinner than the line. Inferior conjunction (IC), the preceding neap position (PN), the following neap position (FN), and Opposition

(O) are labeled.

In addition to the peaks and troughs of M7+ earthquake activity that have already been addressed and which most likely result from Earth passing through a critical angle relative to other planets, Venus and Mars also display two peaks on either side of opposition (e.g., A and B on Fig. 15). These striking and unusual peaks correspond to SOT angles of between 6 and 12 degrees.  In these locations, any of these planets have a magnified potential to slightly perturb the angle of the primary axes of tidal stresses as produced by the sun and moon on Earth; this would in turn, alter the angle of the maximum shear stress, potentially triggering earthquakes either "early" or "late" relative to a Venus or Jupiter that happened to be close to the critical angle. For the M7+ earthquakes shown in the Venus B peak, post-opposition Jupiter SOT angles of 20-30 degrees are over-represented in the data by a factor of 2 to 3 times that which is predicted by random chance, which may indicate that the presence of Venus near opposition is triggering these earthquakes early, before SOT-Jupiter reaches 35 degrees.

Acknowledgements:  Thank you to those individuals that engaged with the authors to discuss early versions of this manuscript: Stephanie Dawson Surberg, Jack Holt, Joann Stock, Tim Melbourne, and John Ashcraft.